\documentclass{elsart}

\usepackage{graphicx,color}

\usepackage{natbib}
\usepackage{amssymb,amsmath,bm}

\newtheorem{theorem}{Theorem}
\newenvironment{proof}{\noindent\textit{Proof}: }{\hfill$\blacksquare$\vskip 0.5\baselineskip}

\newlength\figwidth
\begin{document}

\begin{frontmatter}

\title{On the security of the Yen-Guo's domino signal encryption algorithm
(DSEA)}
\thanks{This paper has been published in \textit{Journal of Systems
and Software}, vol. 79, no. 2, pp. 253-258, 2006.}

\author[cn]{Chengqing Li},
\ead{zjulcq@hotmail.com}
\author[hk]{Shujun Li\corauthref{corr}}
\ead{hooklee@mail.com}
\author[tw]{Der-Chyuan Lou} and
\ead{dclou@ccit.edu.tw}
\author[cnn]{Dan Zhang}
\ead{zhangdan@etang.com}

\address[cn]{Department of Mathematics, Zhejiang University, Hangzhou 310027, China}
\address[hk]{Department of Electronic Engineering, City University of Hong Kong,
83 Tat Chee Avenue, Kowloon Tong, Hong Kong, China}
\address[tw]{Department of Electrical Engineering, Chung
Cheng Institute of Technology, National Defense University,
Taiwan, China}
\address[cnn]{College of Computer Science, Zhejiang University, Hangzhou 310027, China}

\corauth[corr]{The corresponding author, personal web site:
\texttt{http://www.hooklee.com}.}

\begin{abstract}
Recently, a new domino signal encryption algorithm (DSEA) was
proposed for digital signal transmission, especially for digital
images and videos. This paper analyzes the security of DSEA, and
points out the following weaknesses: 1) its security against the
brute-force attack was overestimated; 2) it is not sufficiently
secure against ciphertext-only attacks, and only one ciphertext is
enough to get some information about the plaintext and to break
the value of a sub-key; 3) it is insecure against
known/chosen-plaintext attacks, in the sense that the secret key
can be recovered from a number of continuous bytes of only one
known/chosen plaintext and the corresponding ciphertext.
Experimental results are given to show the performance of the
proposed attacks, and some countermeasures are discussed to
improve DSEA.
\end{abstract}

\end{frontmatter}

\section{Introduction}

In today's networked world, the security issues become more and more
important, so various encryption algorithms have been developed to
fulfill the needs of different applications
\citep{Schneier:AppliedCryptography96}. In recent years, Yen and Guo
et al. proposed a series of chaos-based\footnote{Chaos is a
dynamical phenomenon demonstrated in many dynamical systems
\citep{Devaney:Chaos, HaoBailin:ChaoticDynamics}. Due to the tight
relationship between chaos and cryptography, chaotic systems have
been used to design encryption schemes since 1990s. For a survey of
digital chaotic ciphers, see \citep[Chap. 2]{Li:Dissertation2003}.}
signal/image encryption schemes \citep[Sec.
4.4.3]{Li:ChaosImageVideoEncryption:Handbook2004}, some of which
have been broken according to the works reported in
\citep{ShujunLi:AttackCKBA:ISCAS2002, ShujunLi:AttackBRIE:ICIP2002,
Li:AttackingMES2004, Li:AttackingCNN2004, Li:AttackingRCES2004,
Li:AttackTDCEA2004}. The present paper gives the cryptanalysis
results on a new Yen-Guo encryption scheme called DSEA
\citep{Yen-Guo:DSEA:JCIEE2003}, which has not been cryptanalyzed
before.

DSEA encrypts the plaintext block by block, which is composed of
multiple bytes. The first byte of each block is masked by part of
the secret key, and other bytes are masked by the previous
cipher-byte, under the control of a chaotic pseudo-random bit
sequence (PRBS). That is to say, DSEA works like the dominos. This
paper analyzes the security of DSEA, and points out the following
defects: 1) its security against the brute-force attack was
overestimated; 2) it is not sufficiently secure against
ciphertext-only attacks, and only one ciphertext is enough to get
some information about the plaintext and to break the value of a
sub-key; 3) it is insecure against known/chosen-plaintext attacks,
in the sense that the secret key can be recovered from a number of
continuous bytes of only one known/chosen plaintext and the
corresponding ciphertext.

The rest of this paper is organized as follows. At first,
Sec.~\ref{sec:DSEA} gives a brief introduction to DSEA. Then, the
cryptanalysis results are presented in detail in
Sec.~\ref{sec:Cryptanalysis}, with some experimental results.
Section~\ref{sec:ImprovingDSEA} discusses how to improve DSEA. The
last section concludes the paper.

\section{Domino Signal Encryption Algorithm (DSEA)}
\label{sec:DSEA}

\begin{figure}
\centering
\includegraphics[width=0.7\textwidth]{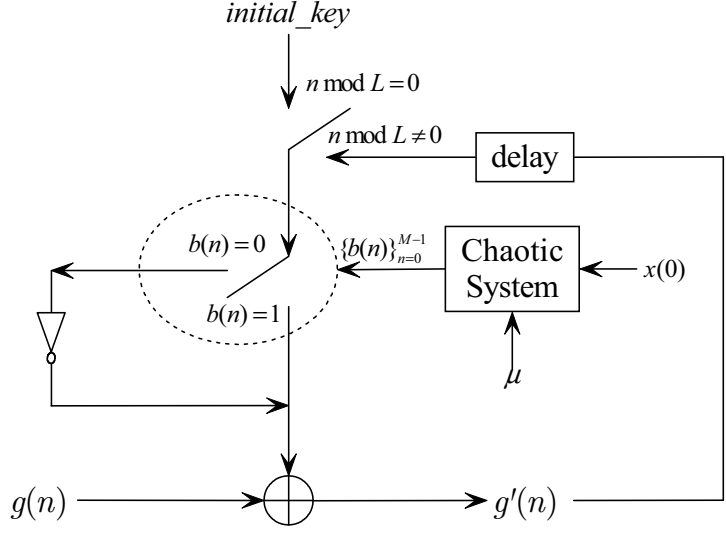}
\caption{The diagrammatic view of the encryption procedure of
DSEA.}\label{figure:DSEA}
\end{figure}

Assume that the plaintext is $g=\{g(n)\}_{n=0}^{M-1}$ and that the
ciphertext is $g'=\{g'(n)\}_{n=0}^{M-1}$, where $g(n)$ and $g'(n)$
denote the $n$-th plain-byte and cipher-byte, respectively. Then,
the encryption procedure of DSEA can be described as follows (see
also Fig.~\ref{figure:DSEA}).

\begin{itemize}
\item\emph{The secret key}: two integers, $L\in\{1,\cdots,M\}$,
$initial\_key\in\{0,\cdots,255\}$, the control parameter $\mu$ and
the initial condition $x(0)$ of the following chaotic Logistic
map\citep{Devaney:Chaos, HaoBailin:ChaoticDynamics}:
\begin{equation}
x(k+1)=\mu\cdot x(k)\cdot(1-x(k)).
\end{equation}

\item\emph{The initialization procedure}: under 8-bit finite
computing precision, run the Logistic map from $x(0)$ to generate a
chaotic sequence $\{x(k)\}_{k=0}^{\lceil M/8\rceil-1}$, and then
extract the 8 significant bits of $x(k)$ to yield a PRBS
$\{b(n)\}_{n=0}^{M-1}$, where $x(k)=\sum_{i=0}^7\left(b_{8k+i}\cdot
2^{-(i+1)}\right)=0.b_{8k+0}\cdots b_{8k+7}$.

\item \emph{The encryption procedure}: for $n=0\sim M-1$, do
\[
g'(n)=
\begin{cases}
g(n)\oplus true\_key, & b(n)=1,\\
g(n)\oplus \overline{true\_key}, & b(n)=0,
\end{cases}
\]
where
\[ true\_key=
\begin{cases}
initial\_key, & n\bmod L=0,\\
g'(n-1), & n\bmod L\neq 0,
\end{cases}
\]
and $\oplus$ denotes the bitwise XOR operation.
\end{itemize}

The decryption procedure is identical with the above encryption
procedure, since XOR is an invertible operation.

\section{Cryptanalysis}
\label{sec:Cryptanalysis}

\subsection{Brute-force attack}

The brute-force attack is the attack of exhaustively searching the
secret key from the set of all possible keys
\citep{Schneier:AppliedCryptography96}. Apparently, the attack
complexity is determined by the size of the key space and the
complexity of verifying each key. The secret key of DSEA is $(L,
initial\_key, \mu, x(0))$, which has $M\cdot 2^{3\cdot 8}=M\cdot
2^{24}$ possible values. Taking the complexity of verifying each key
into consideration, the total complexity of searching for all
possible keys is $O\left(2^{24}\cdot M^2\right)$. When the plaintext
is selected as a typical image of size $256\times 256$, the
complexity will be $O(2^{56})$, which is much smaller than $O(2^M
\cdot M)=O(2^{65552})$, the complexity claimed in
\citep{Yen-Guo:DSEA:JCIEE2003}. Note that the real complexity is
even smaller since not all values of $\mu$ can ensure the chaoticity
of the Logistic map \citep{Devaney:Chaos,
HaoBailin:ChaoticDynamics}. That is, the security of DSEA against
brute-force attacks was over-estimated much in
\citep{Yen-Guo:DSEA:JCIEE2003}. In today's digitized and networked
world, the complexity of order $O(2^{128})$ is required for a
cryptographically-strong cipher
\citep{Schneier:AppliedCryptography96}, which means DSEA is not
practically secure.

\subsection{Ciphertext-only attacks}
\label{sec:CiphertextOnlyAttack}

Ciphertext-only attacks are such attacks in which one can access a
set of ciphertexts \citep{Schneier:AppliedCryptography96}. Since the
transmission channel is generally insecure, the security against
ciphertext-only attacks are required for any ciphers. However, it is
found that DSEA is not sufficiently secure against ciphertext-only
attacks, since much information about the plaintext and the secret
key can be found from even one ciphertext.

Given an observed ciphertext $g'$, generate two mask texts,
$g_0^*$ and $g_1^*$, as follows: $g_0^*(0)=0$, $g_1^*(0)=0,
\forall\; n=1\sim M-1$, $g_0^*(n)=g'(n)\oplus \overline{g'(n-1)}$,
$g_1^*(n)=g'(n)\oplus g'(n-1)$. From the encryption procedure of
DESA, it can be easily verified that the following result is true
when $n \bmod L\neq 0$:
\begin{equation}
g(n)=\begin{cases}
g_0^*(n), & b(n)=0,\\
g_1^*(n), & b(n)=1,
\end{cases}
\end{equation}
which means that $g(n)$ is equal to either $g_0^*(n)$ or
$g_1^*(n)$. Assuming that each chaotic bit distributes uniformly
over $\{0,1\}$, one can deduce that the percentage of right
plain-pixels in $g_0^*$ and $g_1^*$ is not less than
$\frac{L-1}{L}\cdot\frac{1}{2}=\frac{1}{2}-\frac{1}{2L}$. When $L$
is large, about half pixels in $g_0^*$ and $g_1^*$ are
plain-pixels in $g$, and it is expected that some visual
information of the plain-image can be distinguished from $g_0^*$
and $g_1^*$.

\begin{figure}[!htb]
\centering
\begin{minipage}[t]{\figwidth}
\centering
\includegraphics[width=\textwidth]{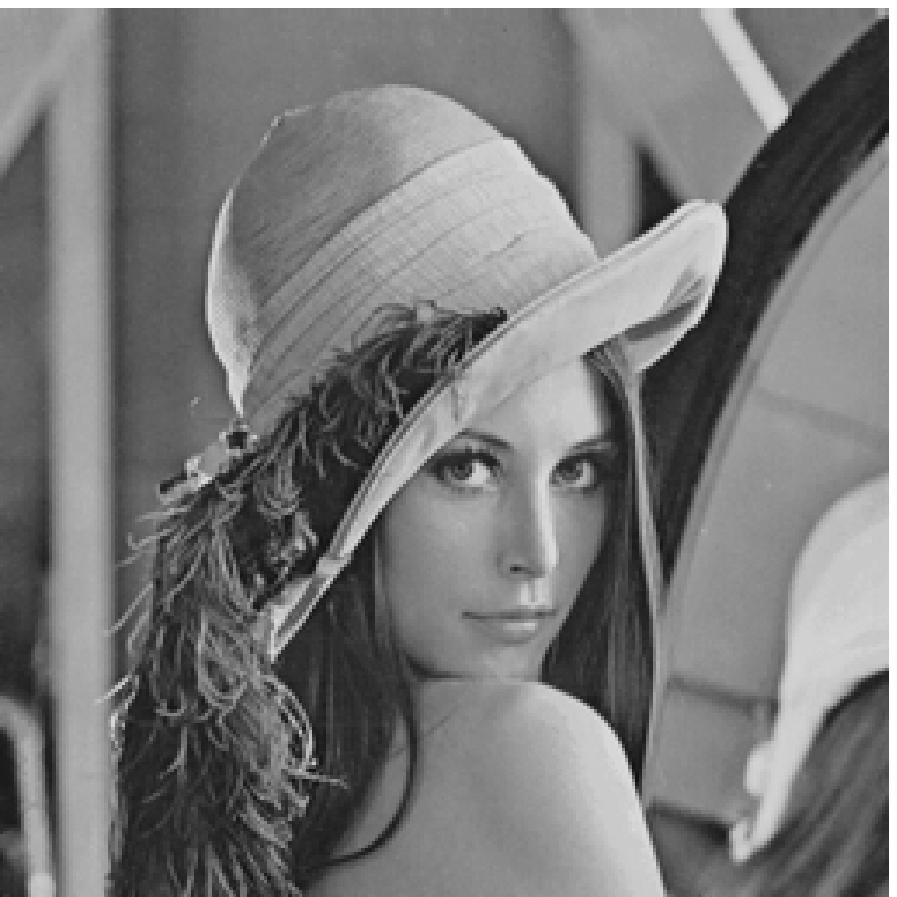}
a) The plain-image $g$
\end{minipage}
\begin{minipage}[t]{\figwidth}
\centering
\includegraphics[width=\textwidth]{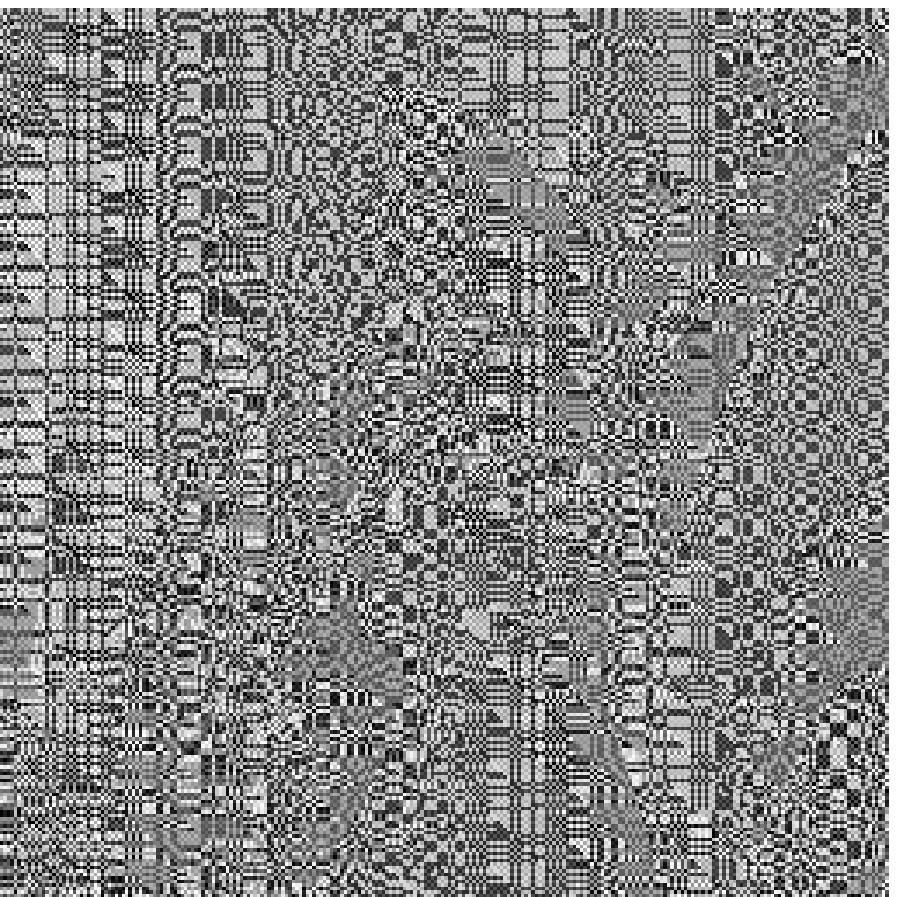}
b) The cipher-image $g'$
\end{minipage}\\
\begin{minipage}[t]{\figwidth}
\centering
\includegraphics[width=\textwidth]{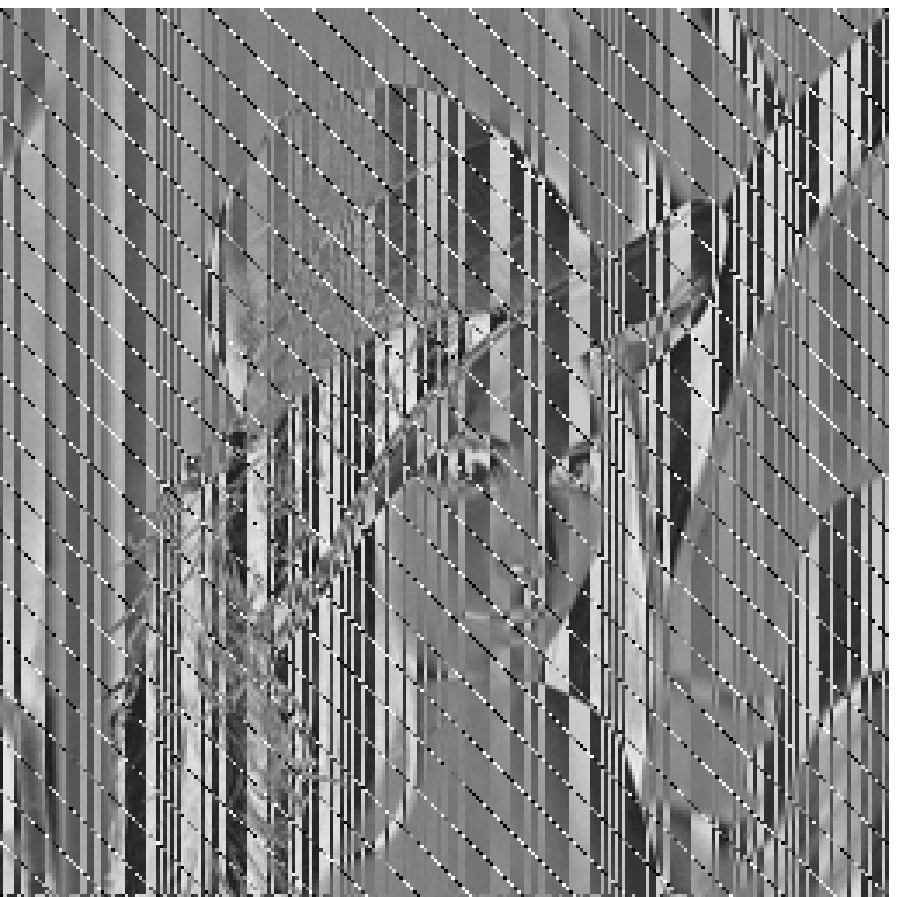}
c) The mask image $g_0^*$
\end{minipage}
\begin{minipage}[t]{\figwidth}
\centering
\includegraphics[width=\textwidth]{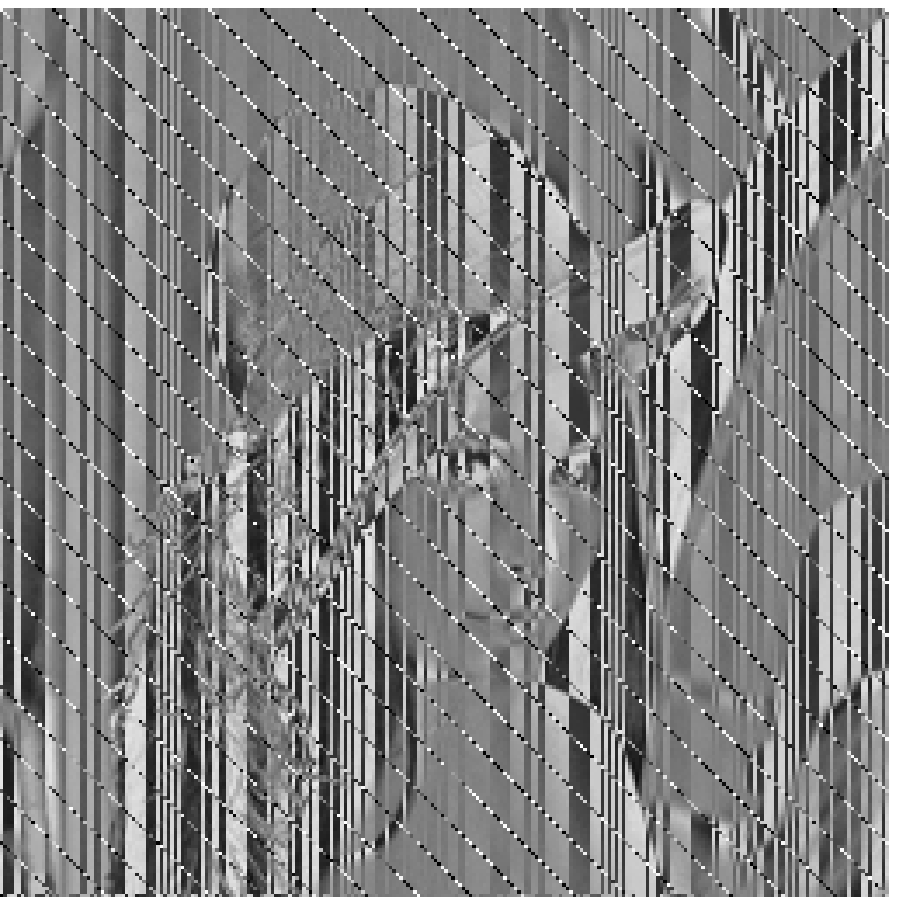}
d) The mask image $g_1^*$
\end{minipage}
\caption{A ciphertext-only attack to DSEA.}
\label{figure:CiphertextOnlyAttack}
\end{figure}

To verify the above idea, one $256\times 256$ image, ``Lenna", has
been encrypted to get $g_0^*$ and $g_1^*$, with the following
secret parameters: $L=15$, $initial\_key=170$, $\mu=251/2^6\approx
3.9219$, $x(0)=69/2^8\approx 0.2695$. The experimental results are
shown in Fig.~\ref{figure:CiphertextOnlyAttack}. In $g_0^*$ there
are 27726 pixels that are identical with those in $g$, and in
$g_1^*$ there are 33461 such pixels. Observing
Figs.~\ref{figure:CiphertextOnlyAttack} c and d, one can see that
the plain-image roughly emerges from both $g_0^*$ and $g_1^*$.

\begin{figure}
\centering
\begin{minipage}[t]{\figwidth}
\centering
\includegraphics[width=\textwidth]{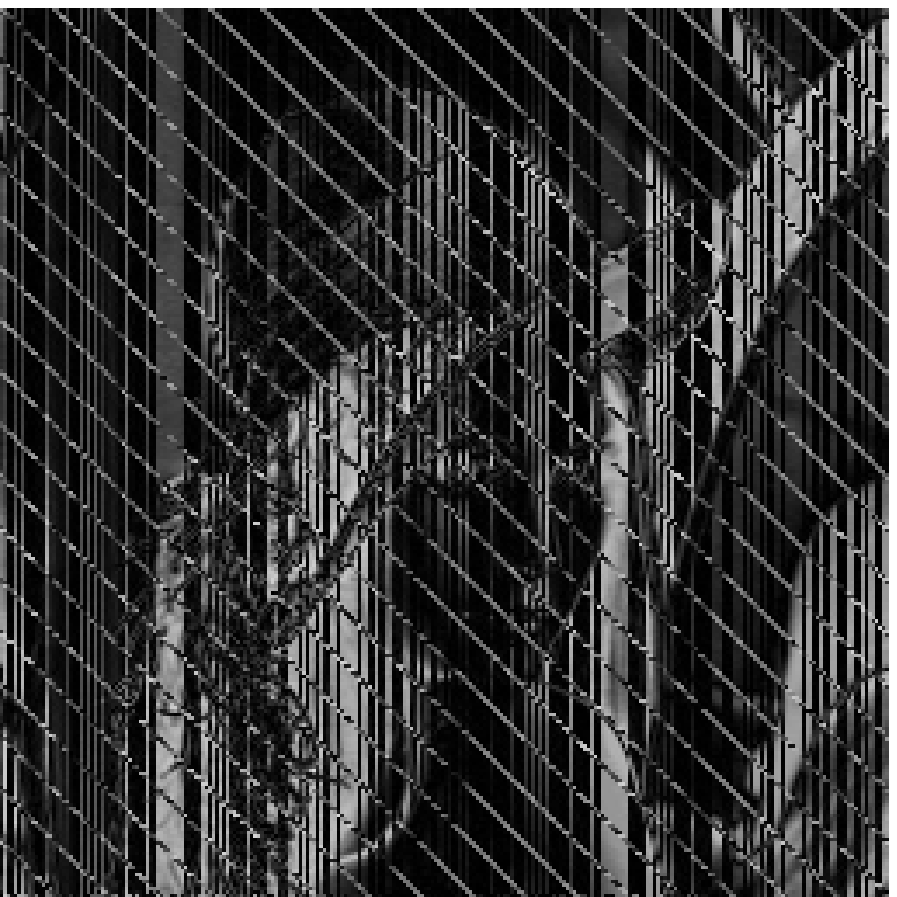}
a) $g_{d,0}^*$
\end{minipage}
\begin{minipage}[t]{\figwidth}
\centering
\includegraphics[width=\textwidth]{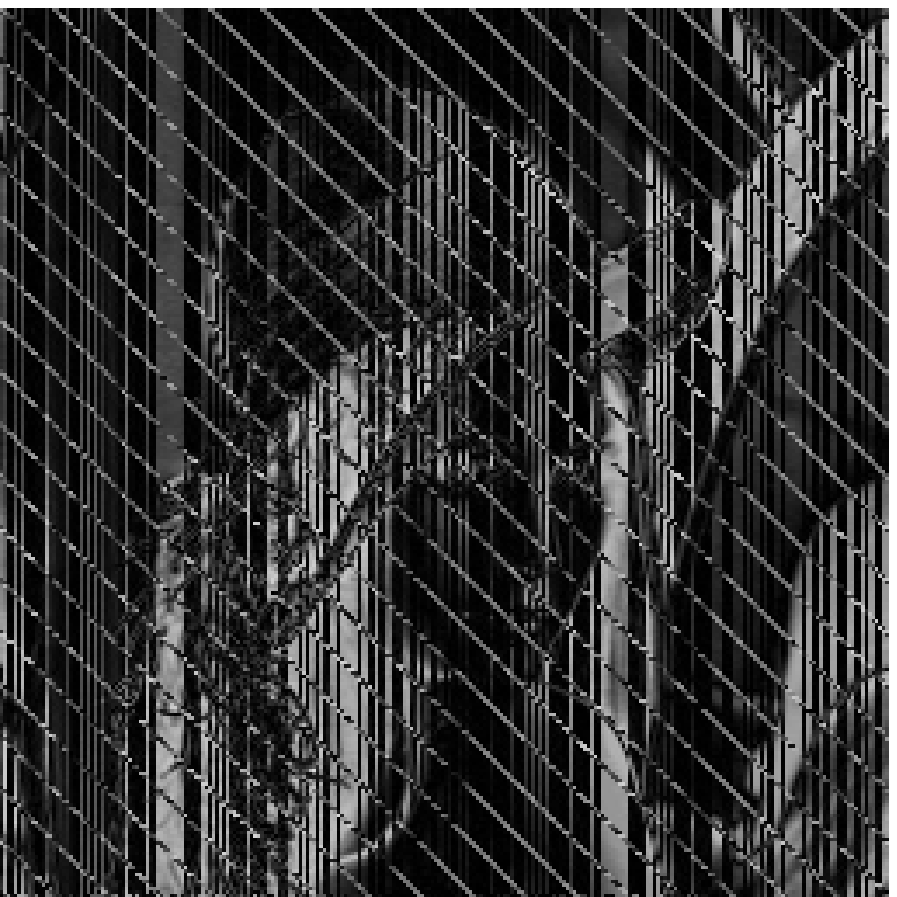}
b) $g_{d,1}^*$
\end{minipage}
\caption{The differential images of $g_0^*$ and $g_1^*$.}
\label{figure:DifferenceImages}
\end{figure}

In addition, from either $g_0^*$ or $g_1^*$, it is possible to
directly get the value of $L$, if there exists strong correlation
between adjacent bytes of the plaintext (speeches and natural
images are good examples). This is due to the probability
difference existing between the following two kinds of
plain-bytes:
\begin{itemize}
\item when $n\bmod L\neq 0$, $g_0^*(n)=g(n)$ and $g_1^*(n)=g(n)$
with a probability of $\frac{1}{2}$;

\item when $n\bmod L=0$, $g_0^*(n)=g(n)$ and $g_1^*(n)=g(n)$ with
a probability\footnote{Without loss of generality, it is assumed
that each cipher-byte distributes uniformly in $\{0,\cdots,255\}$.}
of $\frac{1}{256}$: $g_0^*(n)=g(n)$ if and only if
$g'(n-1)=\overline{initial\_key}$; $g_1^*(n)=g(n)$ if and only if
$g'(n-1)=initial\_key$.
\end{itemize}
When there exists strong correlation between adjacent bytes, the
above probability difference implies that there exists strong
discontinuity around each position satisfying $n\bmod L=0$ (with a
high probability). The fixed occurrence period of such discontinuous
bytes will generate periodically-occurring straight lines in the
mask text when it is an image or displayed in 2-D mode, as shown in
Figs.~\ref{figure:CiphertextOnlyAttack}c and d. Then, it is easy to
determine the occurrence period, i.e., the value of $L$, by checking
the horizontal distance between any two adjacent lines. To make the
straight line clearer, one can calculate the differential images of
$g_0^*$ and $g_1^*$, as shown in Fig.~\ref{figure:DifferenceImages},
where the differential image of an image $g=\{g(n)\}_{n=0}^{M-1}$ is
defined as follows: $g_d(0)=g(0)$ and $\forall\;n=1\sim M-1$,
$g_d(n)=|g(n)-g(n-1)|$. Note that the two differential images of
$g_0^*$ and $g_1^*$ are identical according to the following
theorem, from which one can get that
$|g_0^*(n)-g_0^*(n-1)|=|g'(n)\oplus\overline{g'(n-1)}-g'(n-1)\oplus\overline{g'(n-2)}|
=|g'(n)\oplus g'(n-1)-g'(n-1)\oplus g'(n-2)|=|g_1^*(n)-g_1^*(n-1)|$.

\begin{theorem}
For any three $s$-bit integers, $a,b,c$, it is true that
$|(a\oplus b)-(b\oplus c)|=|(a\oplus\bar{b})-(b\oplus\bar{c})|$.
\end{theorem}
\begin{proof}
Introduce four new variables, $A=a\oplus b$, $B=b\oplus c$,
$A'=a\oplus\bar{b}$, $B'=b\oplus\bar{c}$. It can be easily
verified that $A'=\overline{A}$ and $B'=\overline{B}$, since
$a\oplus\bar{b}=a\oplus b\oplus b\oplus\bar{b}=a\oplus b\oplus
(2^s-1)=\overline{a\oplus b}$. That is, $(a\oplus b)-(b\oplus c)=A-B$
and $(a\oplus\bar{b})-(b\oplus\bar{c})=\overline{A}-\overline{B}$.
Let $A=(A_0\cdots A_{s-1})_2=\sum_{i=0}^{s-1}A_i\cdot 2^i$,
$B=(B_0\cdots B_{s-1})_2=\sum_{i=0}^{s-1}B_i\cdot 2^i$. Since
$\forall\; A_i,B_i\in\{0,1\}$, $A_i-B_i=\bar{B_i}-\bar{A_i}$, it
is obvious that $A-B=\sum_{i=0}^{s-1}(A_i-B_i)\cdot
2^i=\sum_{i=0}^{s-1}(\bar{B_i}-\bar{A_i})\cdot
2^i=\overline{B}-\overline{A}$. As a result, $|(a\oplus
b)-(b\oplus
c)|=|A-B|=|\overline{B}-\overline{A}|=|\overline{A}-\overline{B}|=|(a\oplus\bar{b})-(b\oplus\bar{c})|$,
which completes the proof.
\end{proof}

\subsection{Known/chosen-plaintext attacks}
\label{subsec:KnownPlaintextAttacks}

Known/chosen-plaintext attacks are such attacks in which one can
access/choose a set of plaintexts and observe the corresponding
ciphertexts \citep{Schneier:AppliedCryptography96}. In today's
networked world, such attacks occur more and more frequently. For a
cipher with a high level of security, the security against both
known-plaintext and chosen-plaintext attacks are required. Although
it was claimed that DSEA can resist this kind of attacks \citep[Sec.
IV.B]{Yen-Guo:DSEA:JCIEE2003}, we found this claim is not true: with
a limited number of continuous plain-bytes of only one known/chosen
plaintext, one can completely break the secret key to decrypt other
unknown plain-bytes of the known/chosen plaintext and any new
ciphertexts encrypted with the same key. Apparently, even when the
secret key is changed for each plaintext (as mentioned in
\citep[Sec. IV.B]{Yen-Guo:DSEA:JCIEE2003}), DSEA is insecure against
known/chosen-plaintext attacks. In the following, let us discuss how
to break the four sub-keys, respectively.

\begin{figure}
\centering
\includegraphics[width=\figwidth]{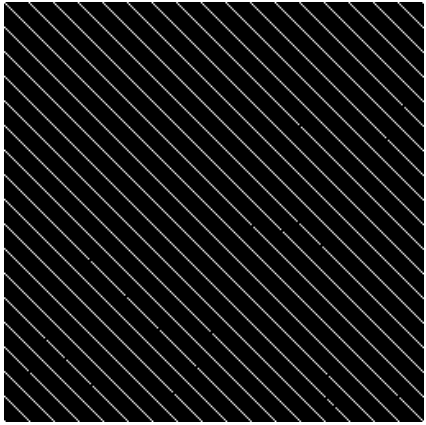}
\caption{The enhanced differential image $g_d^*$.}
\label{figure:EnhancedDifferenceImage}
\end{figure}

\textit{1) Breaking the sub-key $L$:} as mentioned above, once one
gets a ciphertext, he can easily deduce the value of $L$ by
observing the periodically-occurring straight lines in the two
constructed mask texts, $g_0^*$ and $g_1^*$. Furthermore, since
the plaintext is also known, it is possible to generate an
enhanced differential image, $g_d^*$, as follows: $g_d^*(0)=0$,
and $\forall\; n=1\sim M-1$,
\begin{equation}
g_d^*(n)=\begin{cases} 0, & g(n)\in\{g_0^*(n), g_1^*(n)\},\\
255, & g(n)\not\in\{g_0^*(n), g_1^*(n)\}.
\end{cases}
\end{equation}
See Fig.~\ref{figure:EnhancedDifferenceImage} for the enhanced
differential image corresponding the cipher-image shown in
Fig.~\ref{figure:CiphertextOnlyAttack}b. Compared with
Fig.~\ref{figure:DifferenceImages}, one can see that the straight
lines become clearer.

\textit{2) Breaking the $initial\_key$:} for all values of $n$
that satisfy $n\bmod L=0$, it is obvious that
\begin{equation}\label{equation:GetInitialKey}
initial\_key=
\begin{cases}
g(n)\oplus g'(n), & b(n)=1,\\
\overline{g(n)\oplus g'(n)}, & b(n)=0.
\end{cases}
\end{equation}

Note that it is possible to uniquely determine the value of
$initial\_key$, when there may exist pixels satisfying $n\bmod
L=0$ and $g_d^*(n)=0$, i.e., $g(n)\in\{g_0^*(n),
g_1^*(n)\}=\left\{g'(n)\oplus \overline{g'(n-1)},g'(n)\oplus
g'(n-1)\right\}$. Considering $g'(n)=g(n)\oplus initial\_key$, one
can immediately deduce that
\begin{equation}\label{equation:GetInitialKey}
initial\_key=
\begin{cases}
g'(n-1), & g(n)=g_1^*(n),\\
\overline{g'(n-1)}, & g(n)=g_0^*(n).
\end{cases}
\end{equation}

\textit{3) Breaking the chaotic PRBS and the other two sub-keys:}
once $L$ and $initial\_key$ have been determined, the chaotic
PRBS, $\{b(n)\}_{n=0}^{M-1}$, can be immediately derived as
follows:
\begin{itemize}
\item when $n \bmod L\neq 0$: if $g(n)=g_0^*(n)$ then $b(n)=0$,
else $b(n)=1$;

\item when $n \bmod L=0$: if $initial\_key=g(n)\oplus g'(n)$ then
$b(n)=1$, else $b(n)=0$.
\end{itemize}

Once $\{b(n)\}_{n=0}^{M-1}$ is uniquely determined, $x(0)=0.b(0)\cdots b(7)$
can be immediately recovered.

With 16 consecutive chaotic bits, $b(8k+0)\sim b(8k+15)$, one can
further derive two consecutive chaotic states:
$x(k)=0.b(8k+0)\cdots b(8k+7)$ and $x(k+1)=0.b(8k+8)\cdots
b(8k+15)$, and then derive an estimation of the sub-key $\mu$ as
\begin{equation}
\widetilde{\mu}=\frac{x(k+1)}{x(k)\cdot(1-x(k))}.
\end{equation}
Due to the quantization errors introduced in the finite-precision
arithmetic, generally $x(k+1)\neq\mu\cdot x(k)\cdot(1-x(k))$, so
$\widetilde{\mu}\neq\mu$. Fortunately, following the error analysis
of $\widetilde{\mu}$ in \citep[Sec. 3.2]{Li:AttackingCNN2004}, the
following result has been obtained: when $x(k+1)\geq 2^{-n}\;(n=1\sim
8)$, $|\widetilde{\mu}-\mu|<2^{n+3}\cdot 2^{-8}$. Specially, when
$x(k+1)\geq 2^{-1}=0.5$, $|\widetilde{\mu}-\mu|<2^4\cdot 2^{-8}$,
which means that one can exhaustively search for $2^4=16$ values in
the neighborhood of $\widetilde{\mu}$ to find the right value of
$\mu$. To verify which searched value is the right one, one can
iterate the Logistic map from $x(k+1)$ for some times to get some
new chaotic states and then check the coincidence between these
chaotic states and corresponding recovered chaotic bits.

\begin{figure}[!htb]
\centering
\includegraphics[width=\figwidth]{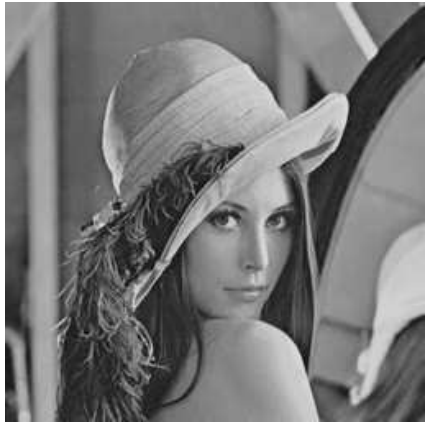}
\caption{The recovered plain-image of ``Lenna" in a
known-plaintext attack.} \label{figure:KnownPlaintextAttack}
\end{figure}

With the above steps, the whole secret key
$(L,initial\_key,\mu,x(0))$ can be recovered, and then be used for
decryption. For the plain-image ``Lenna", a breaking result is
shown in Fig.~\ref{figure:KnownPlaintextAttack}. It can be
verified that the complexity of the known/chosen-plaintext attacks
is only $O(M)$, which means a perfect breaking of DSEA.

\section{Improving DSEA}
\label{sec:ImprovingDSEA}

In this section, we study some possible remedies to DSEA to resist
the proposed attacks. It is concluded that DSEA cannot be simply
enhanced to resist known/chosen-plaintext attacks.

To ensure the complexity of the brute-force attack
cryptographically large, the simplest idea is to increase the
presentation precision of $x(0)$ and $\mu$. Binary presentations
of $x(0)$ and $\mu$ with 64-bit (long integers) are suggested to
provide a complexity not less than $O(2^{128})$ against the
brute-force attack.

Apparently, the insecurity of DSEA against ciphertext-only and
known/chosen-plaintext attacks is mainly due to the invertibility of
XOR operations. This is actually the weakness of all XOR-based
stream ciphers. To make DSEA securer, one has to change the
encryption structure and/or the basic masking operations, in other
words, one has to design a completely new cipher, instead of
enhancing DSEA to design a modified cipher.

\begin{figure}[!htb]
\centering
\begin{minipage}[t]{\figwidth}
\centering
\includegraphics[width=\figwidth]{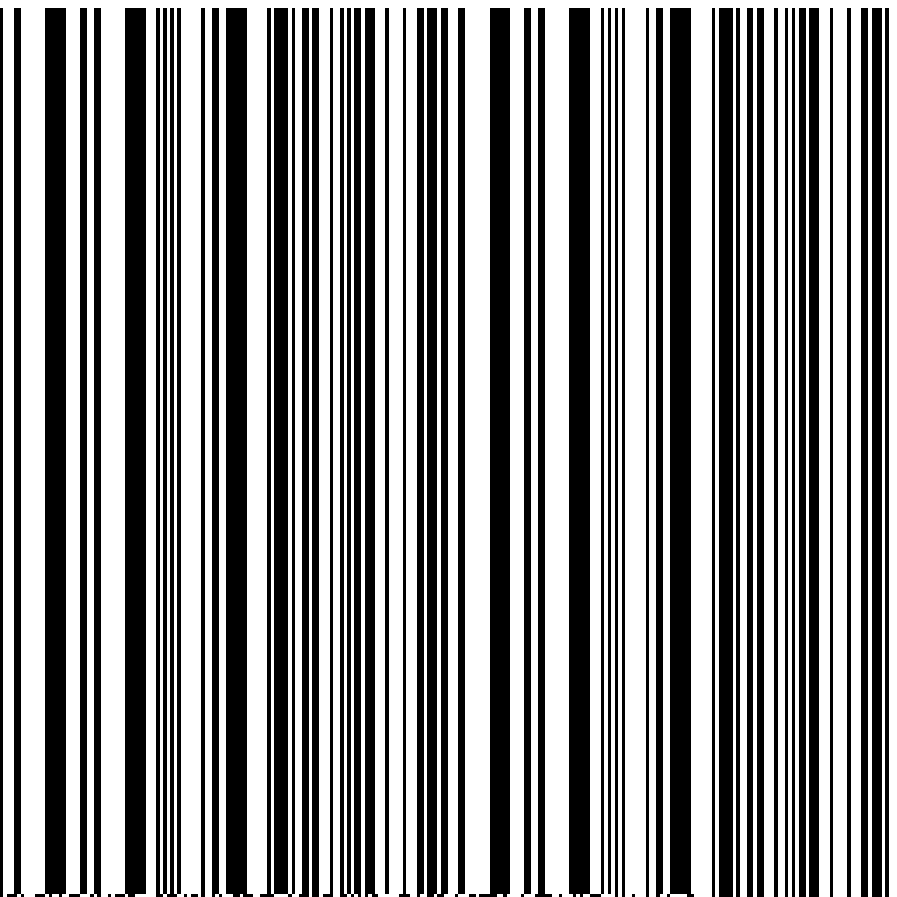}
a) 8-bit fixed-point arithmetic
\end{minipage}
\begin{minipage}[t]{\figwidth}
\centering
\includegraphics[width=\figwidth]{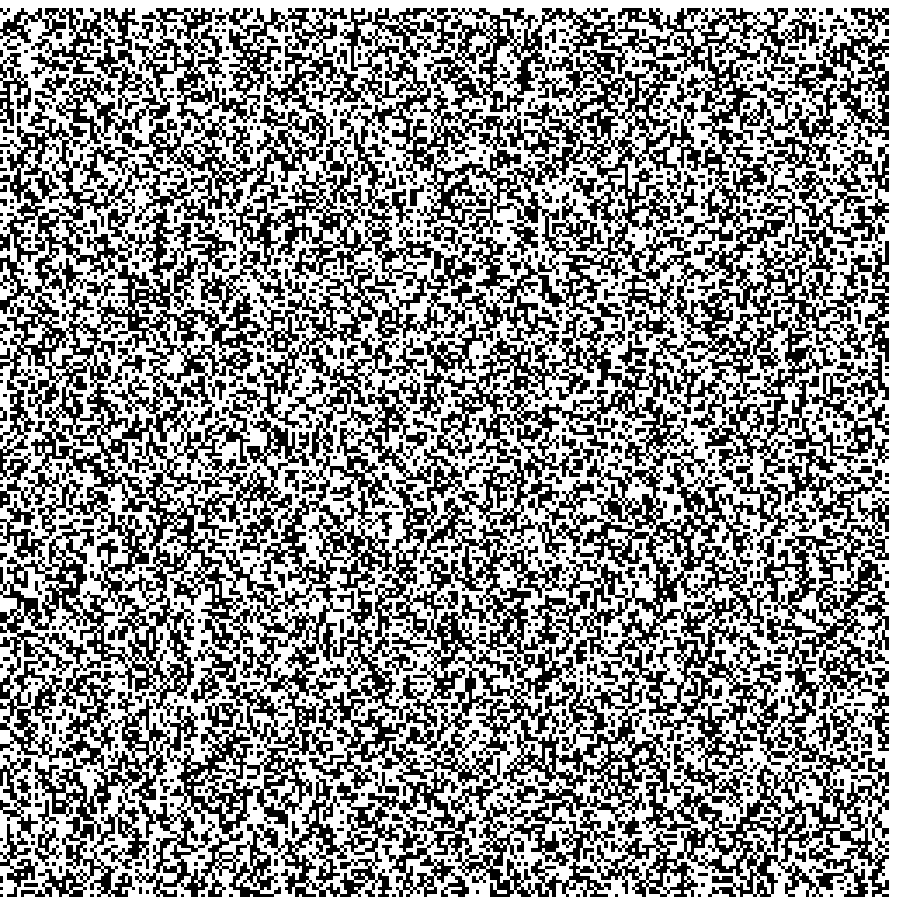}
b) double-precision floating-point arithmetic
\end{minipage}
\caption{The pseudo-image of the chaotic PRBS, under two different
finite-precision arithmetics.} \label{figure:ChaoticBits}
\end{figure}

In addition, there exists a special flaw in DSEA. According to
\citep[Sec. 2.5]{Li:Dissertation2003}, when a chaotic system is
implemented in $s$-bit finite computing precision, each chaotic
orbit will lead to a cycle whose length is smaller than $2^s$ (and
generally much smaller than $2^s$). Figure~\ref{figure:ChaoticBits}a
shows the pseudo-image of the chaotic PRBS recovered in a
known-plaintext attack. It is found that the cycle of the chaotic
PRBS is only $2^6=64$ and the period of the corresponding chaotic
orbit is only $2^3=8$. Such a small period of the chaotic PRBS will
make all attacks easier. To amend this defect, using a higher
implementation precision or floating-point arithmetic is suggested.
Figure~\ref{figure:ChaoticBits}b gives the pseudo-image of the
chaotic PRBS when the chaotic states are calculated under
double-precision floating-point arithmetic. It is obvious that the
short-period effect of the chaotic PRBS is effectively avoided.

\section{Conclusion}

In this paper, the security of a recently-proposed signal security
system called DSEA \citep{Yen-Guo:DSEA:JCIEE2003} has been studied
in detail. It is pointed out that DSEA is not secure enough against
the following attacks: the brute-force attack, ciphertext-only
attacks, and known/chosen-plaintext attacks. Experimental results
are also given to support the theoretical analysis. Also, some
remedies of enhancing the performance of DSEA are discussed. As a
conclusion, DSEA is not suggested in serious applications requiring
a high level of security.

\section{Acknowledgements}

This research was partially supported by the National Natural
Science Foundation, China, under grant no. 60202002, and by the
Applied R\&D Centers of the City University of Hong Kong under
grants nos. 9410011 and 9620004.

\bibliographystyle{elsart-harv}
\bibliography{DSEA}

\end{document}